\documentclass[letterpaper,twocolumn]{jpsj3}
\usepackage{txfonts}
\usepackage{color}

\title{
Magnetic-Field Dependence of Paramagnetic Properties Investigated by $^{63/65}$Cu-NMR on the Yb Zigzag-Chain Semiconductor YbCuS$_2$
}

\author{Fumiya Hori$^1$\thanks{hori.fumiya.36s@st.kyoto-u.ac.jp}, Shunsaku Kitagawa$^1$, Kenji Ishida$^1$,  Yudai Ohmagari$^2$ and Takahiro Onimaru$^2$}
\inst{$^1$Depertment of Physics, Kyoto University, Kyoto 606-8502, Japan \\
$^2$Department of Quantum Matter, Graduate School of Advanced Science and Engineering, Hiroshima University,
Higashihiroshima 739-8530, Japan \\
} 

\abst{
To investigate the paramagnetic properties of YbCuS$_2$ under magnetic fields, we have performed the $^{63/65}$Cu-nuclear magnetic resonance (NMR) measurements.
The NMR spectra can be reproduced by the simulations of the three-dimensional powder pattern and the additional two-dimensional powder pattern, indicating the partial sample orientation due to the anisotropy of the magnetic properties.
These simulations suggest that the $ac$ plane is the easy plane in YbCuS$_2$.
The Knight shift $K$ is proportional to the bulk magnetic susceptibility and field-independent.
The broad maximum of the nuclear spin-lattice relaxation rate $1/T_1$ at $T_{\rm max} \sim 50$~K (50~K anomaly) observed at zero magnetic field is quickly suppressed by the magnetic fields.
This indicates that the 50~K anomaly is field-dependent.
Furthermore, an anomalous enhancement of $1/T_1$ at low temperatures was observed above 3~T.
This field seemingly corresponds to the magnetic field at which a field-induced phase transition occurs below the antiferromagnetic transition temperature $T_{\rm N} \sim 1$~K.
The changes in $1/T_1$ observed in the paramagnetic state suggest the presence of the complex quantum phenomena under magnetic fields in YbCuS$_2$.
}

\begin{document}
\maketitle

\section{Introduction}
Recently, there has been significant attention on magnetic frustration in insulating or semiconducting compounds based on rare-earth ions from Ce ($4f^1$) to Yb($4f^{13}$)~\cite{Yb_frustration, Saito1, YbMgGaO4-1, YbMgGaO4-2, YbMgGaO4-3, NaRSe2, NaYbSe2, NaYbSe2_spinon_fermi_surface, Yb_triangular_1, Yb_triangular_2, Yb_triangular_3, Ohmagari1, Ohmagari2, Hori2022, Hori2023, Hori_YbAgSe2, Saito2, Saito3}.
The interplay of the strong spin–orbit coupling and the crystalline electric field (CEF) effect of $4f$ electrons leads to the anisotropic exchange interactions, leading to the presence of exotic magnetic states not observed in standard quantum magnets based on transition metals~\cite{Yb_frustration, Saito1}.
Indeed, Yb-based insulators YbMgGaO$_4$~\cite{YbMgGaO4-1, YbMgGaO4-2, YbMgGaO4-3} and NaYbSe$_2$~\cite{NaRSe2, NaYbSe2, NaYbSe2_spinon_fermi_surface}, with Yb$^{3+}$ triangular lattices, have been proposed to exhibit novel ground states, such as $Z_2$ spin liquids and spinon Fermi surfaces~\cite{ Yb_triangular_1, Yb_triangular_2, Yb_triangular_3}.

\begin{figure*}[t]
\centering
\includegraphics[width=10cm]{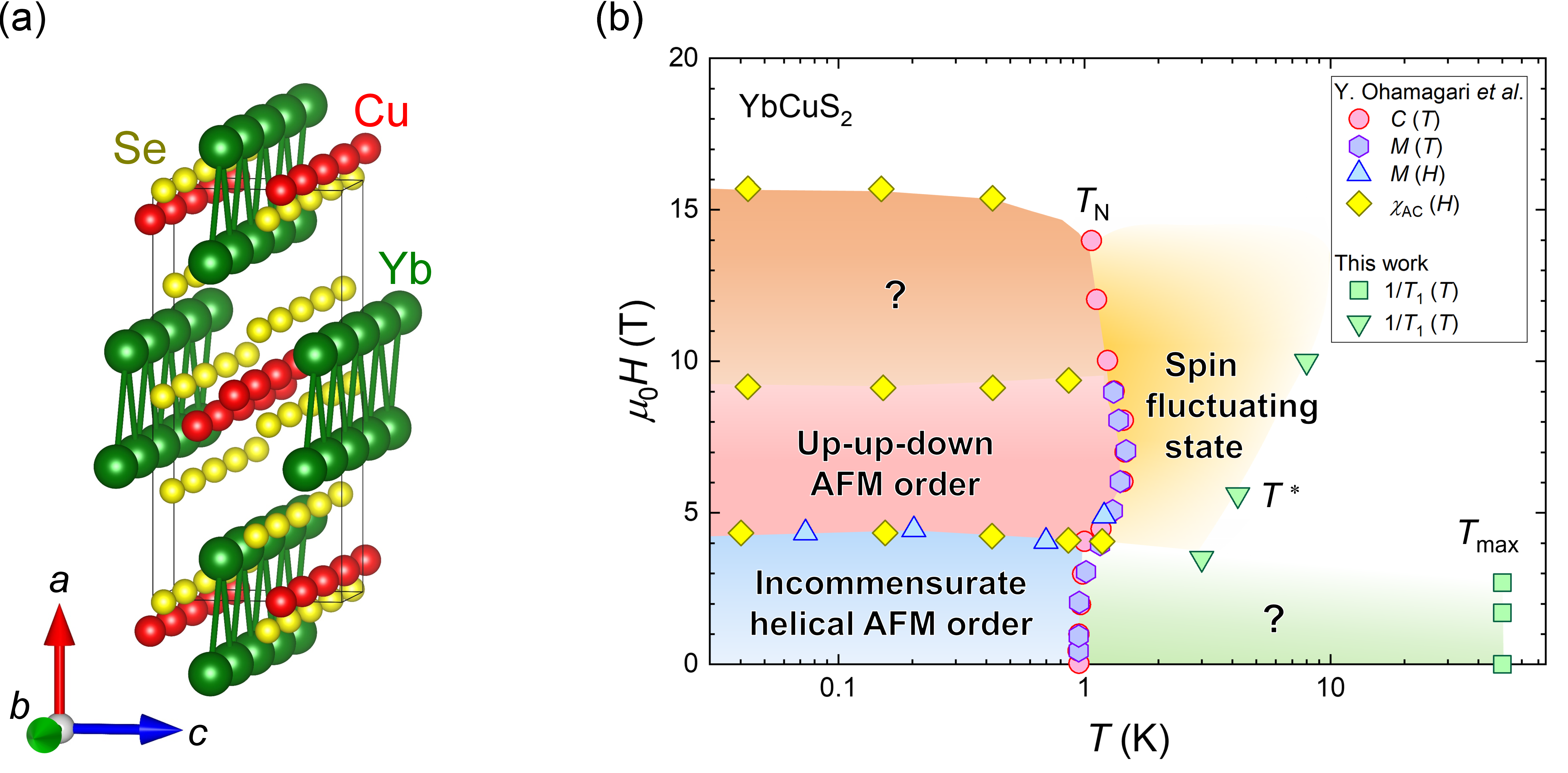} 
\caption{(Color online) (a) Crystal structure of YbCuS$_2$. The solid line represents the unit cell. 
(b) 
$H$-$T$ phase diagram of YbCuS$_2$; the circles and hexagons are the transition temperatures $T_{\rm N}$ obtained from the highest peaks of the specific heat $C(T)$ data and the kinks of the magnetization $M(T)$ data, respectively.~\cite{Ohmagari2}
The ordered phase is divided into at least three phases as indicated by the triangles given by the peaks of $dM(H)/dH$ and the rhombuses obtained by the anomalies of $\chi_{\rm AC} (H)$.~\cite{Ohmagari2}
The squares are $T_{\rm max}$ obtained from the maximum of $1/T_1$ below 3~T.
The inverted triangles are $T^{\ast}$, below which $1/T_1$ deviates from the extrapolation line below 30~K.}
\label{structure}
\end{figure*}

We have focused on a Yb-based frustrated semiconductor
YbCuS$_2$ crystallizing in an orthorhombic structure.
This system has Yb$^{3+}$ zigzag chains as shown in Fig.~\ref{structure}(a), where the frustration effect driven by the competition between the nearest-neighbor exchange interactions $J_1$ and the next-nearest-neighbor exchange interactions $J_2$ is expected~\cite{Saito1, Saito2, Saito3}.
Several unique properties resulting from this frustration effect induced by the Yb zigzag chains 
were consistently observed~\cite{Ohmagari1, Ohmagari2, Hori2022, Hori2023}.
At around $T_{\rm max} \sim 50$~K, the magnetic entropy $S_{\rm m}$ reaches $R \ln 2$ ($R$ is the gas constant), and 
the $^{63/65}$Cu-nuclear quadrupole resonance (NQR) nuclear spin-lattice relaxation rate $1/T_1$ shows a broad maximum (50~K anomaly).
The origin of the 50~K anomaly remains unclear.
This compound shows a first-order magnetic transition at $T_{\rm N} \sim 0.95$~K characterized by the strong divergence of the specific heat and the coexistence of the paramagnetic and antiferromagnetic (AFM) NQR signals near $T_{\rm N}$.
$S_{\rm m}$ at $T_{\rm N}$ remains only 20\% of $R \ln 2$ expected for the Kramers doublet CEF ground state.
Below $T_{\rm N}$, an incommensurate helical structure is confirmed by the recent neutron diffraction experiments~\cite{Onimaru_YbCuS2_neutron} and $^{63/65}$Cu-NQR measurements~\cite{Hori2023}.
The ordered moment 0.2-0.4~$\mu_{\rm B}$ is much smaller than the anticipated value 1.3~$\mu_{\rm B}$ from
the doublet CEF ground state.
The $T$-linear behavior of $1/T_1$ was observed on the AFM state, indicating the presence of the gapless quasiperticle excitations~\cite{Hori2023}.
The transition temperature $T_{\rm N} \sim 0.95$~K at zero magnetic field is much lower than 
the paramagnetic Curie temperature
 $\theta_{\rm p} \sim -32$~K,
reflecting the  presence of the frustration effect.

In addition, under the magnetic field $H$, YbCuS$_2$ also exhibits intriguing properties, which are not seen in conventional antiferromagnets~\cite{Ohmagari2}.
The magnetic field vs temperature ($H$-$T$) phase diagram of
YbCuS$_2$ is presented in Fig.~\ref{structure}(b).
The transition temperature $T_{\rm N}$ is robust against magnetic fields for $\mu_0 H < 4$~T, and $T_{\rm N}(H)$ increases to higher temperatures for 4~T~$<\mu_0 H < 7$~T.
When a magnetic field $\mu_0 H > 4$~T is applied, the magnetization shows a 1/3 plateau.
In the 1/3 plateau phase, a commensurate up-up-down configuration is theoretically expected~\cite{Saito3, zigzag1, zigzag3}, reflecting the magnetic frustration effect.
Furthermore, for $\mu_0 H > 9$~T, another distinct magnetic phase emerges, suggesting the presence of complex quantum phenomena in this system under magnetic fields~\cite{Ohmagari2}.

In this paper, we report the $^{63/65}$Cu-nuclear magnetic resonance (NMR) results to investigate the paramagnetic properties of YbCuS$_2$ under magnetic fields from a microscopic perspective.
The NMR spectra can be reproduced by the simulations of the three-dimensional (3D) powder pattern and the additional two-dimensional (2D) powder pattern corresponding to the partial sample orientation by applying $H$.
These simulations suggest that the $ac$ plane is the easy plane.
The Knight shift $K$ is proportional to the bulk magnetic susceptibility and $H$-independent.
The 50~K anomaly of $1/T_1$ is suppressed by the magnetic fields.
The low-temperature fluctuations are enhanced by applying $H$, and the upturn of $1/T_1$ was observed above 3~T.
This behavior seems to be related to the $H$-induced phase transition below $T_{\rm N}$.
The changes in $1/T_1$ suggest that the $H$-induced modification occurs even in the paramagnetic phase.

\section{Experimental}

Polycrystalline samples of YbCuS$_2$ were synthesized by the melt-growth method~\cite{Ohmagari1, Ohmagari2}.
The polycrystalline samples were coarsely powdered to increase the surface area for better thermal contact.
A conventional spin-echo technique was used for the  $^{63/65}$Cu-NMR measurements.
$^{63}$Cu and $^{65}$Cu nuclei with spin $I = 3/2$ have nuclear gyromagnetic ratios of $^{63}\gamma/2\pi = 11.289$~MHz/T and $^{65}\gamma/2\pi =12.093$~MHz/T, respectively.
In order to determine the quadrupolar and magnetic Knight shift separately, we took series of $H$-swept $^{63/65}$Cu-NMR spectra at several different frequencies in the range of 19.5~MHz (1.7~T)- 113~MHz (10~T).
The temperature dependence of the Knight shift was estimated at the central-transition peak with the highest intensity as indicated with arrows in Fig.~\ref{Knight_shift}(a).
The $^{63}$Cu-NMR $1/T_1$ was also measured at the peak with the highest intensity.
The $^{63}$Cu-NMR $1/T_1$ was evaluated by fitting the relaxation curve of the nuclear magnetization after the saturation to a theoretical function for the nuclear spin $I = 3/2$.
In whole $H$ region, $T_1$ was evaluated from the reliable fitting of the relaxation curve with the single component of $T_1$ as shown later.

\section{Results and Discussion}

\begin{figure}[t]
\centering
\includegraphics[width=7cm]{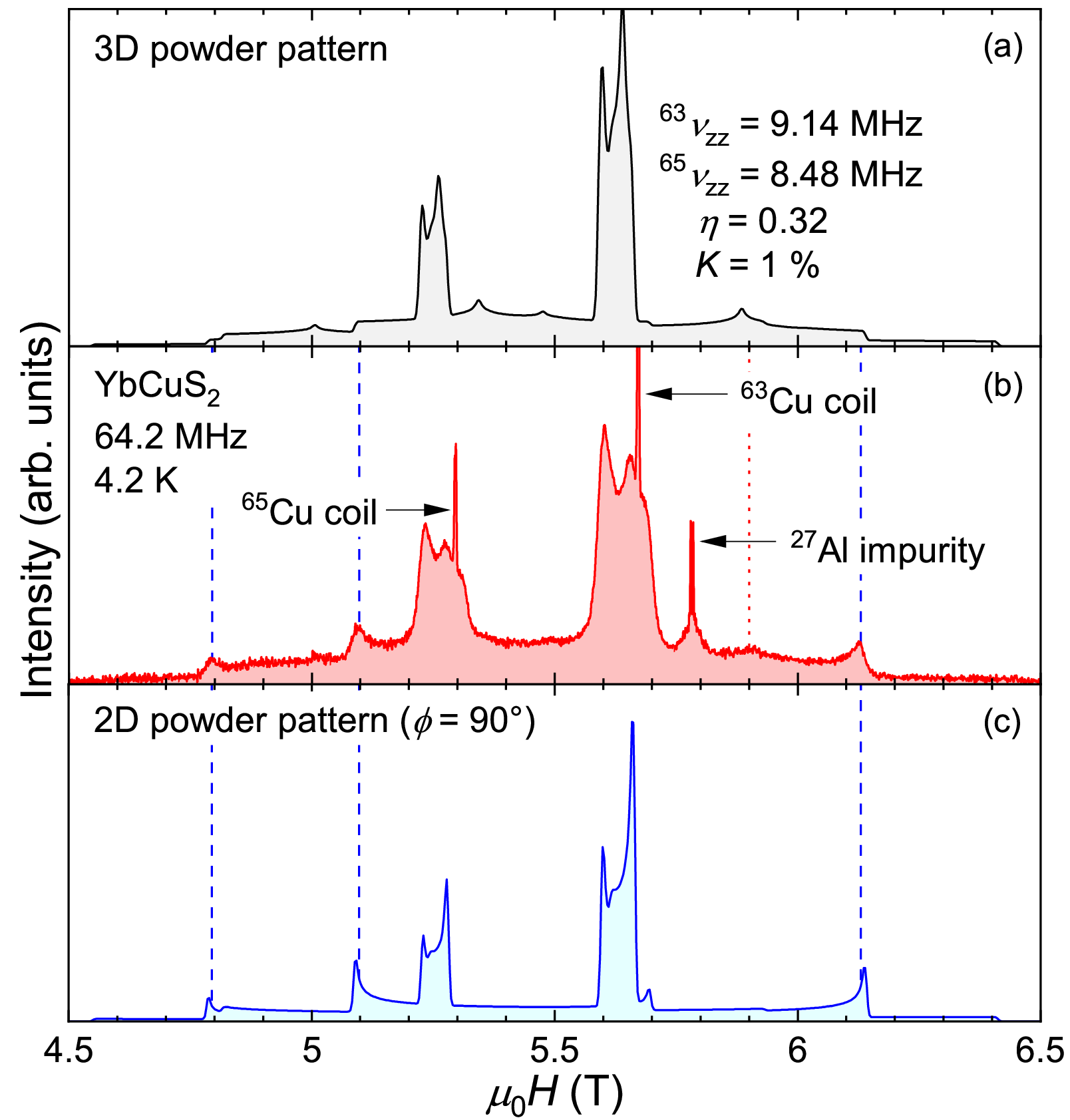} 
\caption{(Color online) (a) 3D powder pattern simulation with $K = 1$\%, $^{63}\nu_{zz} =$ 9.14~MHz, $^{65}\nu_{zz} =$ 8.48~MHz, and $\eta = 0.32$.
(b) Observed NMR spectrum at a fixed frequency of 64.2~MHz and 4.2~K in YbCuS$_2$; the narrow lines in the NMR spectrum are signals from the Cu coil used for the NMR measurements and Al as an impurity arising from the NMR probe.
(c) 2D powder pattern simulation fixing the azimuthal angle of $ \phi = 90^\circ$ ($H \perp V_{xx}$) with $K = 1$\%, $^{63}\nu_{zz} =$ 9.14~MHz, $^{65}\nu_{zz} =$ 8.48~MHz, and $\eta = 0.32$; 
the blue dashed lines indicate the oriented peaks, and 
the red dotted line indicates the peak which can explained by the 3D powder pattern simulation.}
\label{simulation}
\end{figure}

\begin{figure*}[t!]
\centering
\includegraphics[width=16cm]{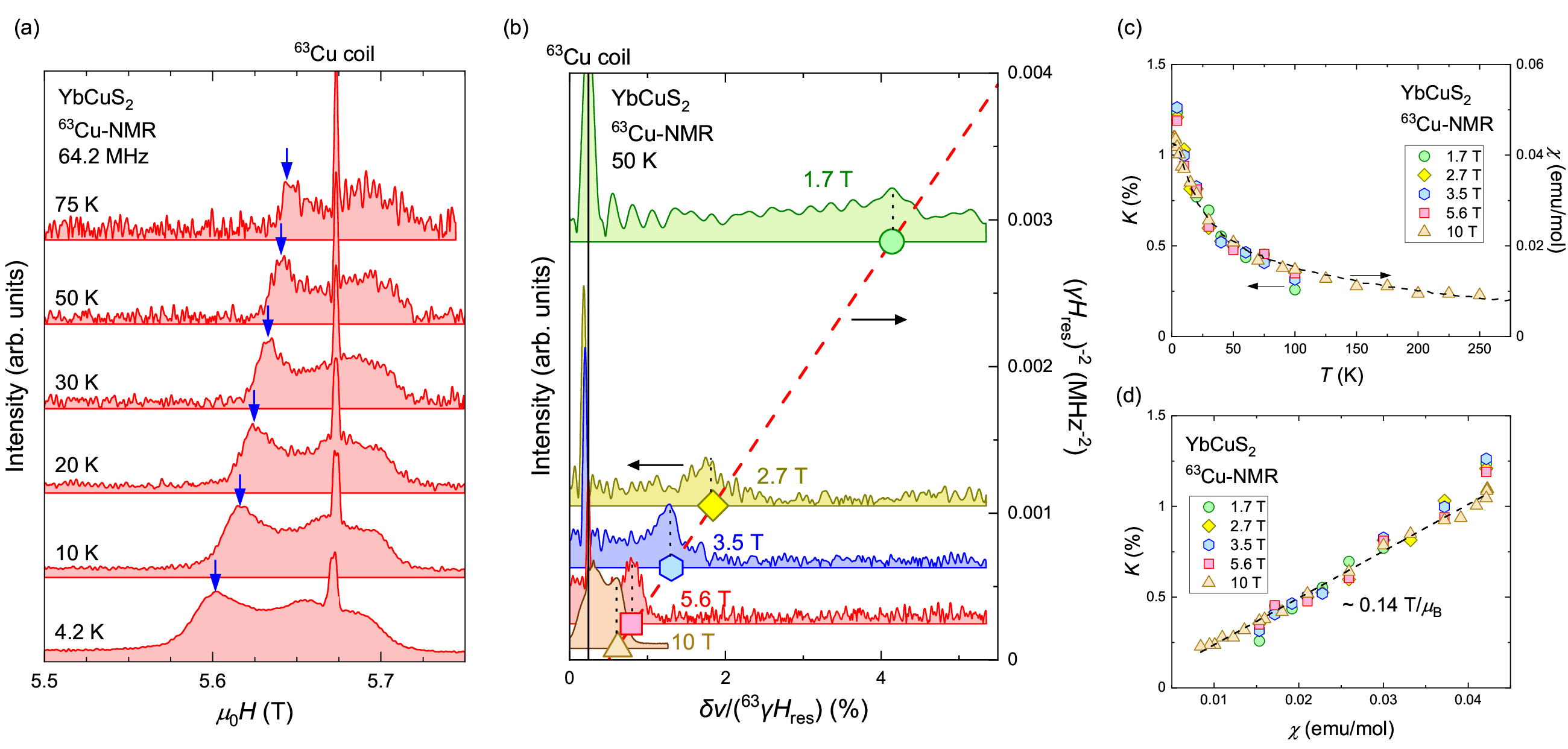} 
\caption{(Color online) (a) Temperature variations of the $^{63}$Cu-NMR spectrum of powdered YbCuS$_2$ obtained by $H$-swept method at a frequency of 64.2~MHz for 4.2~K $ \leq T \leq$~75~K; the temperature dependence of the Knight shift was estimated at the peak with the highest intensity as indicated in the arrows.
(b) Field variations of the $^{63}$Cu-NMR spectrum (left hand) and $(^{63}\gamma H_{\mathrm{res}})^{-2}$
for different NMR frequency (right hand); the horizontal axis is $\delta v/^{63}\gamma H_{\mathrm{res}}$.
The circles, rhombuses, hexagons, squares, and triangles represent 1.7~T (19.5~MHz), 2.7~T (31.4~MHz), 3.5~T (40.4~MHz), 5.6~T (64.2~MHz) and 10~T (113~MHz), respectively.
(c) Temperature dependence of $^{63}$Cu-NMR Knight shift $K$ and bulk magnetic susceptibility $\chi$. (d) $K$-$\chi$ plot with the temperature as an implicit parameter.
}
\label{Knight_shift}
\end{figure*}

Figures~\ref{simulation}(b) shows the $H$-swept $^{63/65}$Cu-NMR spectrum measured at the frequency of $\nu = 64.2$~MHz on the powdered sample.
In general, the total effective NMR Hamiltonian of a nucleus in $H$ is given by
\begin{flalign}
\mathcal{H} &=\mathcal{H}_{\mathrm{Z}}+\mathcal{H}_{\mathrm{Q}}, \nonumber \\
\mathcal{H}_{\mathrm{Z}}&= -\frac{\gamma}{2 \pi} h (1+K) \boldsymbol{I} \cdot \boldsymbol{H}, \nonumber \\
\mathcal{H}_{\mathrm{Q}}&=
\frac{h \nu_{z z}}{6}\left\{\left(3 I_{z}^{2}-I^{2}\right)+\frac{1}{2} \eta\left(I_{+}^{2}+I_{-}^{2}\right)\right\},
\end{flalign}
where $\gamma$ is the nuclear gyromagnetic ratio, $K$ is the Knight shift, $h$ is the Planck constant, and $I_{\pm}$ are the ladder operators of the nuclear spin $I$, which are defined as $I_{\pm} = I_x \pm iI_y$.
$ \nu_{zz}~(\propto V_{zz})$ is the quadrupole frequency along the principal axis of the electric field gradient (EFG) and is defined as $h \nu_{zz} \equiv 3eV_{zz}Q/2I(2I-1)$ with the electric quadrupole moment $Q$. $\eta$ is the asymmetry parameter of the EFG defined as $\eta \equiv (V_{xx} -V_{yy} )/V_{zz}$, where $V_{ii}$ is the second derivative of the electric potential $V$, $V_{ii} = \partial ^2 V/\partial x_{i}^2 $.
Since the NMR signal of each grain depends on the angle between the principal axis of the EFG $V_{zz}$ in each grain and the direction of $H$, the sum of the NMR signals for all 3D solid angles could be observed in the powder samples.
The observed NMR spectra were partially reproduced by the 3D powder pattern simulation with $K = 1$\%, $^{63}\nu_{zz} =$ 9.14~MHz, $^{65}\nu_{zz} =$ 8.48~MHz, and $\eta = 0.32$, as shown in Fig.~\ref{simulation}(a).
However, some peaks indicated by the blue dashed lines could not be explained by this simulation, which are attributed to partial orientation of the sample due to the magnetic anisotropy. 
The oriented peaks were consistently reproduced by the 2D powder pattern, calculated as the sum of NMR signals over all angles in the 2D plane fixing the azimuthal angle of $ \phi = 90^\circ$ ($H \perp V_{xx}$), as shown in Fig.~\ref{simulation}(c).
According to WIEN2k calculation using the density functional theory~\cite{WIEN2k}, $V_{zz}$ is almost parallel to the $b$ axis in YbCuS$_2$.
Thus, the observed partial orientation can be parallel to the $ac$ plane.
In fact, the $ac$-plane rotating incommensurate helical magnetic structure at $H = 0$ below $T_{\rm N}$ was confirmed by our NQR results~\cite{Hori_NQR_pressure} and neutron  diffraction measurements~\cite{Onimaru_YbCuS2_neutron}.
Therefore, it is reasonable to consider that the $ac$ plane and the $b$ axis are the easy plane and the hard axis in the AFM state of YbCuS$_2$, which is consistent with the observed orientation in the paramagnetic state.

Figure~\ref{Knight_shift}(a) shows the temperature variations of the $^{63/65}$Cu-NMR spectrum.
As indicated by the arrows, the peak with the highest intensity shifts toward lower $H$ as the temperature decreases.
For nuclei with $I=1/2$, the Knight shift is given by $\delta v/\gamma H_{\mathrm{res}}$, where $ H_{\rm res}$ is the magnetic field at which the resonance peak is observed, and $\delta \nu = \nu - \gamma H_{\rm res}$.
However, for nuclei with $I>1/2$ as in the case of this study, it is not so simple due to the presence of quadrupole parameters $\nu_{zz}$ and $\eta$.
In this case, the Knight shift can be evaluated from a series of spectra at several different frequencies $\nu$.
In this method as described below, the above-mentioned simulation using the whole of the powder pattern is not necessary for estimating the Knight shift.
We can express $\delta \nu$ for $\mathcal{H}_{\mathrm{Z}} \gg \mathcal{H}_{\mathrm{Q}}$ as~\cite{Takigawa_plot, Takigawa_plot2}
\begin{flalign}
\begin{aligned}
\frac{\delta v (T)}{^{63}\gamma H_{\mathrm{res}}(T)} \sim K(T)+\frac{D(T)}{(1+K(T))\left(^{63}\gamma H_{\mathrm{res}}(T)\right)^2}.
\end{aligned}
\label{D-K}
\end{flalign}
The first term in the right side represents the Knight shift, and the second term $D$ is the second-order effect of the quadrupolar interaction.
Since Eq.~(\ref{D-K}) is a linear relation between $\delta v/^{63}\gamma H_{\mathrm{res}}$ and $(^{63}\gamma H_{\mathrm{res}})^{-2}$,
$K$ and $D$ can be estimated from the intercept and slope, respectively.
$D \sim 12.81$~(MHz)$^2$ at 50~K was estimated as shown in the dashed line of Fig.~\ref{Knight_shift}(b).
We assumed that $D$ is $T$-independent, and $K$ is given by
\begin{flalign}
\begin{aligned}
K(T) \sim \frac{\delta v (T)}{^{63}\gamma H_{\mathrm{res}}(T)} - \frac{D({\rm 50 K})}{(1+K(T))\left(^{63}\gamma H_{\mathrm{res}}(T)\right)^2}.
\end{aligned}
\end{flalign}
Figure~\ref{Knight_shift}(c) exhibits the temperature dependence of $K$ estimated at several different frequencies.
$K$ was correctly estimated in the above-mentioned process.
It was found that $K$ is $H$-independent and increases on cooling.
Moreover, $K$ follows the Curie-Weiss behavior, which is scaled with the bulk magnetic susceptibility $\chi$.
The hyperfine coupling constant $A_{\rm{hf}}$ is the proportionality constant between the Knight shift and magnetic susceptibility, 
\begin{flalign}
\begin{aligned}
K(T) = A_{\rm{hf}} \chi (T) + K_{0}.
\end{aligned}
\end{flalign}
Thus, $A_{\rm{hf}}$ can be obtained from the slope of the $K$-$\chi$ plot.
Here, $K_{0}$ is the $T$-independent term.
From the $K$-$\chi$ plot with $\chi$ measured on a polycrystalline sample at 0.1~T~\cite{Ohmagari1, Ohmagari2}, $A_{\rm{hf}}$ was estimated to be 0.14~T/$\mu_{\rm{B}}$ and remains constant on cooling down to the lowest temperature as shown in Fig.~\ref{Knight_shift}(d).

In general, a hyperfine coupling constant in insulators or semiconductors originates from the classical dipolar interaction, which contributes to the anisotropic term in the Knight shift and is given by~\cite{Hori_YbAgSe2}
\begin{flalign}
\begin{aligned}
A_{\rm d}&=\sum_{n} 9.2741 \times 10^{-1} \\
&\times \frac{1}{r_n^{5}}\left(\begin{array}{ccc}
3 x_{n}^{2}-r_n^{2} & 3 x_n y_n & 3 x_n z_n \\
3 y_n x_n & 3 y_i^{2}-r_n^{2} & 3 y_n z_n \\
3 z_n x_n & 3 z_n y_n & 3 z_n^{2}-r_n^{2}
\end{array}\right) \;\; \left({\rm T}/ \mu_{B}\right),
\label{dipole}
\end{aligned}
\end{flalign}
where $r_n = \sqrt{x_n^2 + y_n^2 + z_n^2}$~($\AA$) is a distance between Cu nuclei and Yb-$n$ sites, and $x_n$, $y_n$ and $z_n$ are the components parallel to the $a$, $b$, and $c$ axes in YbCuS$_2$, respectively.
From Eq. (\ref{dipole}),
\begin{flalign}
{A}_{\rm d}= \left(\begin{array}{ccc}
-0.005 & 0.000 &  -0.011 \\
0.000 & -0.035& 0.000 \\
-0.011 & 0.000 & 0.041
\end{array}\right) \;\; \left({\rm T} / \mu_{B}\right)
\end{flalign}
is obtained.
The experimental value of $A_{\rm{hf}} = 0.14$~T/$\mu_{\rm{B}}$ is larger than all components of $A_{\rm{d}}$.
Thus, contributions except for the classical dipolar interactions may exist.
This mainly originates from the transferred hyperfine field from the magnetic Yb site to the NMR atomic site~\cite{CeB6_NMR_transferred, LiCuVO4_NMR, Taniguchi_2016_PrTi2Al20}.
In the previous study of the $^{77}$Se-NMR measurements of YbAgSe$_2$~\cite{Hori_YbAgSe2}, which is a sister compound of YbCuS$_2$, the Knight shift is negative.
In both cases, the hyperfine coupling constant is larger than the estimated values of the classical dipolar interaction, and the sign of the coupling in YbAgSe$_2$ is negative due to the core polarization of Se 4$p$~\cite{Hori_YbAgSe2}.
To estimate accurately the value of $A_{\rm hf}$, the NMR measurements on a single crystal are required.

\begin{figure}[t]
\centering
\includegraphics[width=9cm]{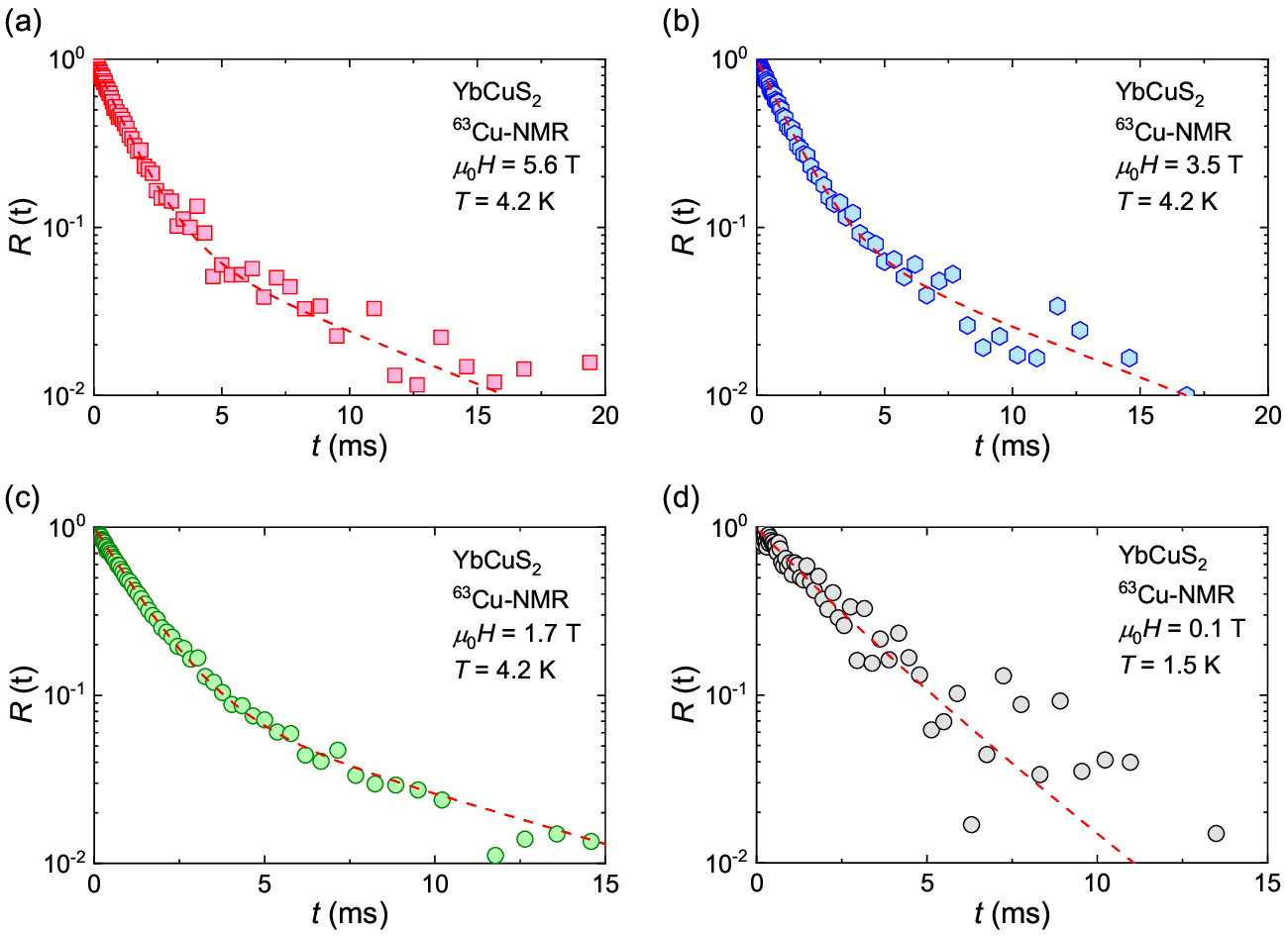} 
\caption{(Color online) $^{63}$Cu relaxation curve $R(t)$ at (a) $\mu_0H = 5.6$~T, $T = 4.2$~K, (b) $\mu_0H = 3.5$~T, $T = 4.2$~K, (c) $\mu_0H = 1.7$~T, $T = 4.2$~K, and (d) $\mu_0H = 0.1$~T, $T = 1.5$~K.
The dashed line represents the fitting with the single component of $T_1$.
}
\label{relaxation_curve}
\end{figure}

\begin{figure}[t]
\centering
\includegraphics[width=8cm]{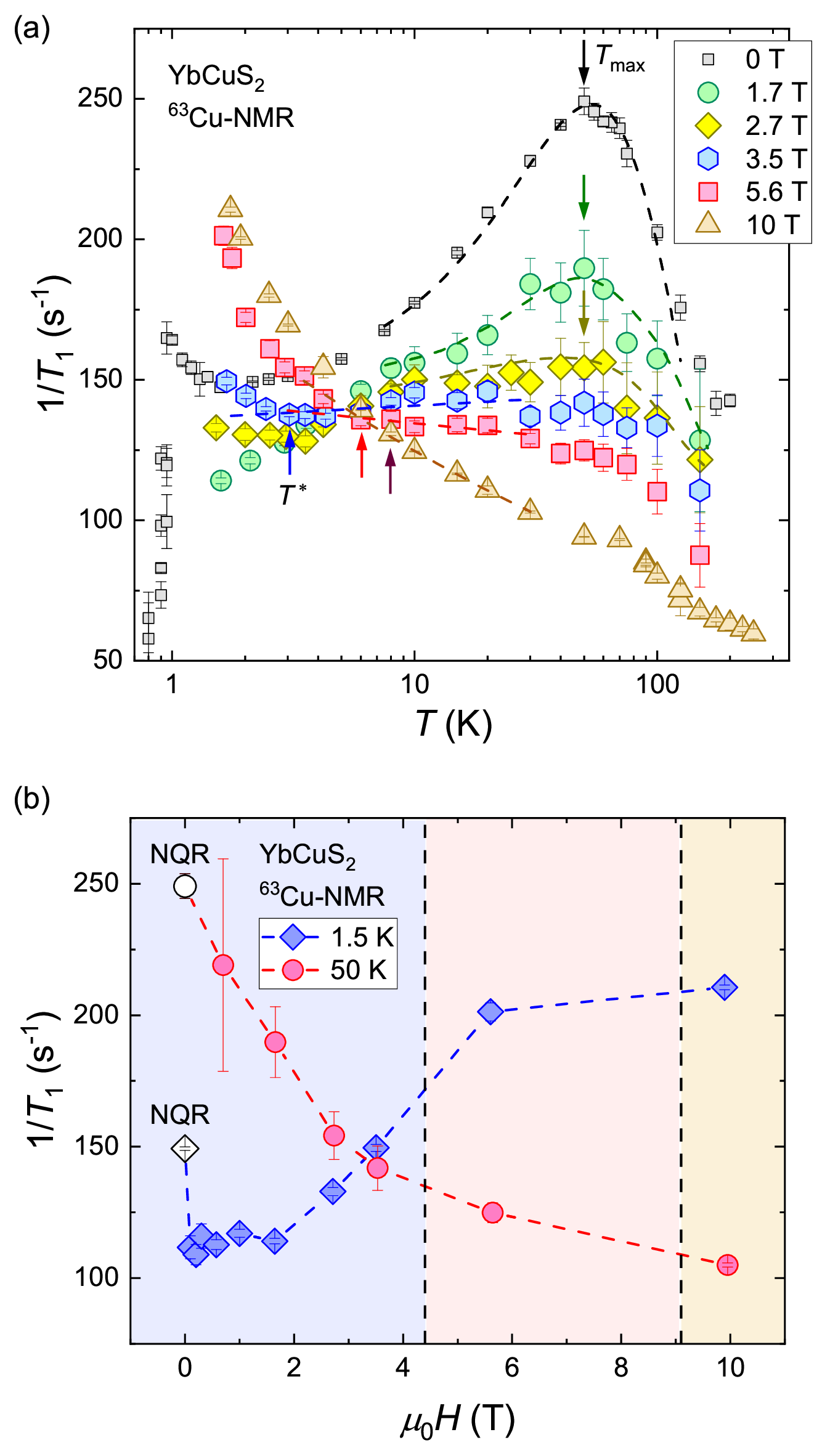} 
\caption{(Color online) (a) Temperature dependence of $^{63}$Cu-NMR $1/T_1$ on YbCuS$_2$ at several magnetic fields;
the downarrows represent $T_{\rm max}$ obtained from the maximum of $1/T_1$ below 3~T.
The uparrows indicates $T^{\ast}$, below which $1/T_1$ deviates from the extrapolation line below 30~K.
The dashed curves are guides for eyes.
(b) Magnetic field dependence of $^{63}$Cu-NMR $1/T_1$ on YbCuS$_2$; circles and diamonds represent $1/T_1$ at 50~K and 1.5~K, respectively. The dashed vertical lines represent the field-induced magnetic transitions below $T_{\rm N}$.
}
\label{1_T1}
\end{figure}

Next, we focus on $T$ and $H$ dependence of low-energy spin dynamics probed with the $1/T_1$ measurements.
Figure~\ref{relaxation_curve} shows the typical relaxation curves $R(t) \equiv 1-M(t) / M (\infty)$ of the nuclear magnetization $M(t)$ and the fitting for $T_1$ determination.
As shown in Fig.~\ref{relaxation_curve}(a)-(c), a theoretical function for the central-transition line of $I = 3/2$,
\begin{flalign}
R(t) = 0.1 \exp \left(-\frac{ t}{T_1}\right)
+0.9 \exp \left(-\frac{ 6t}{T_1}\right)
\end{flalign}
was adopted as the fitting function for $\mu_0 H > 0.5$~T.
In low-$H$ region $\mu_0 H < 0.5$~T, we used the theoretical relaxation curves obtained by the matrix diagonalization~\cite{general_relaxation_curve} on the assumption that $H \parallel V_{yy}$, $K = 1$\%, $^{63}\nu_{zz} =$ 9.14~MHz, and $\eta = 0.32$.
For example, in the case of $\mu_0 H = 0.1$~T, 
\begin{flalign}
\begin{aligned}
R(t) = 0.123 \exp \left(-\frac{ 2.714t}{T_1}\right)
+0.563 \exp \left(-\frac{ 3.727t}{T_1}\right) \\
+0.314\exp \left(-\frac{ 5.959t}{T_1}\right)
\end{aligned}
\end{flalign}
was adopted as the fitting function as shown in Fig~\ref{relaxation_curve}(d).
In whole $H$ region, $T_1$ was evaluated from the reliable fitting of the relaxation curve with the single component of $T_1$.

Figures~\ref{1_T1}(a) and \ref{1_T1}(b) show the $T$ and $H$ dependence of the $^{63}$Cu-NMR $1/T_1$, respectively.
As previously reported~\cite{Hori2022, Hori2023}, the zero-field $^{63}$Cu-NQR $1/T_1$ exhibits a broad maximum at $T_{\rm max} \sim 50$~K (50~K anomaly).
With increasing $H$, the $1/T_1$ value at 50~K decreases as shown in Fig.~\ref{1_T1}(b), indicating that the 50~K anomaly is suppressed by $H$.
Above 3~T, the clear broad maximum was not observed.
Note that the 50~K anomaly remains unchanged under pressure or chemical substitutions at $H = 0$ as previously reported~\cite{Hori_NQR_pressure, Hori_NQR_substitution}, while it is strongly affected by $H$.

Focusing on the low-temperature region, $1/T_1$ decreases on cooling down to 1.5~K below $T_{\rm max}$ in low magnetic fields of $\mu_0 H < 2$~T.
However, for $\mu_0 H > 3$~T, the slope of $1/T_1$ changes at low temperatures, and the upturn of $1/T_1$ was observed.
We define $T^{\ast}$ as the temperature below which $1/T_1$ deviates from the extrapolation line below 30~K.
$T^{\ast}$ increases with $H$.
At 1.5~K, just above $T_{\rm N}$, the $1/T_1$ value increase with $H$. 
This $H$ dependence at 1.5~K contrasts with that at 50~K, as shown in Fig.~\ref{1_T1}(b).
Note that discontinuous change in $1/T_1$ at 1.5~K was observed across zero and finite magnetic fields,
due to the difference of the magnetic fluctuations detectable with the measurements.
The NQR $1/T_1$ at $H = 0$ probes the fluctuations perpendicular to $V_{zz}$, whereas $1/T_1$ under finite $H$ probes fluctuations perpendicular to the $H$ direction, and thus the fluctuations parallel to $V_{zz}$ are also probed.

$T_{\rm max}$ and $T^{\ast}$ determined in this study are presented in the $H$-$T$ phase diagram of Fig.~\ref{structure}(b). 
For $\mu_{0} H < 3$~T, the 50~K anomaly at $T_{\rm max}$ was observed. 
The 50~K anomaly becomes invisible and
a spin fluctuating state is realized below $T^{\ast}$ for $\mu_{0} H > 3$~T.
This behavior indicates the magnetically driven modification of the paramagnetic phases.
This magnetically driven modification above $T_{\rm N}$ seems to correspond to the $H$-induced magnetic phase transition below $T_{\rm N}$.
We speculate that the 50~K anomaly is linked to the incommensurate helical AFM ordered state below $T_{\rm N}$, while the spin fluctuating state is related to the up-up-down AFM ordered state at low $T$.

Here, we discuss the origin and characteristics of the 50~K anomaly.  
A similar broad maximum of $1/T_1$ has been observed in other Yb-based frustrated magnets~\cite{NaYbSe2, NaYbO2_NMR, NaYbSe2_NMR}.  
In these compounds, the development of the short-range magnetic correlations at low temperatures or the CEF effects have been proposed~\cite{NaYbO2_NMR, NaYbSe2_NMR}. 
Below the temperature where $1/T_1$ shows the anomaly, a spectrum broadening was detected in such materials.
However, such a spectrum broadening was not detected in YbCuS$_2$.  
In addition, the 50~K anomaly in YbCuS$_2$ is robust against chemical substitutions and pressure, as reported in a previous study~\cite{Hori_NQR_pressure, Hori_NQR_substitution}. 
These results suggest that the origin of the 50~K anomaly in YbCuS$_2$ may not be simply attributed to the development of the short-range magnetic correlations or the CEF effects.
Furthermore, a frequency-dependent maximum of $1/T_1$ can often be understood by Bloembergen–Purcell–Pound model~\cite{BPP_model}, where $1/T_1 \sim \exp(\Delta/T)/[1 + \nu^2 \exp(2\Delta/T)]$, as is commonly applied to some spin-glass systems~\cite{Spin_glass_BPP_model}.
However, the 50~K anomaly in YbCuS$_2$ cannot be fitted by such a model.  
Recently, it has been theoretically proposed that the anisotropic exchange interactions arise due to the Yb zigzag chains, leading to the coexistence of a gapped dimer state and a gapless nematic ordered state~\cite{Saito1, Saito2, Saito3}. 
It seems that the 50~K anomaly might be related to the formation of such non-trivial ordered states, although the detailed origin remains unclear.

Finally, we point out the similarity between the spin fluctuating state in YbCuS$_2$ and that observed in several one-dimensional $S = 1/2$ spin-ladder systems~\cite{Cu2(C5H12N2)2Cl4_NMR, BPCB_NMR}.
In such one-dimensional systems, similar $H$-induced changes in the slope and value of $1/T_1$ have been reported.
Theoretically, a $H$-driven transition from a nonmagnetic gapped state to a Tomonaga–Luttinger liquid state has been proposed~\cite{Cu2(C5H12N2)2Cl4_NMR, BPCB_NMR}.
The spin-fluctuating states in YbCuS$_2$ may originate from such a $H$-induced spin liquid state observed in one-dimensional systems, although the ground state in YbCuS$_2$ is the AFM ordered state.
In the case of the theoretical one-dimensional zigzag-chain model with $J_1 \sim J_2$, which is relevant to YbCuS$_2$, a 1/3 magnetization plateu phase with up-up-down magnetic structures occurs as the ground state under magnetic fields~\cite{Saito3, zigzag1, zigzag3}.
This can explain the magnetization results below $T_{\rm N}$ in YbCuS$_2$, as mentioned in the introduction part.
However, it is unclear whether an experimentally observed spin-fluctuating state appears at finite temperatures under magnetic fields in the theoretical zigzag chain model.
To further investigate the magnetic properties in YbCuS$_2$, NMR measurements below $T_{\rm N}$ under a magnetic field are essential, which are currently in progress.

\section{Conclusion} 

In conclusion, we have performed the $^{63/65}$Cu-NMR measurements to investigate the paramagnetic properties of YbCuS$_2$ under magnetic fields.
The NMR spectra can be reproduced by the simulations of the 3D powder pattern and the additional 2D powder pattern corresponding to the partial sample orientation under magnetic fields.
These simulations indicates that the $ac$ plane is the easy plane in YbCuS$_2$.
The Knight shift $K$ is proportional to the bulk magnetic susceptibility and $H$-independent.
The 50~K anomaly of $1/T_1$ is strongly suppressed by magnetic fields.
This indicates that the 50~K anomaly is $H$-susceptible.
The low-temperature fluctuations are enhanced and the upturn of $1/T_1$ was observed in magnetic fields for $\mu_0 H > 3$~T.
Our results suggest that the $H$-induced modification occurs even in the paramagnetic phase and highlight the realization of complex magnetic phenomena in YbCuS$_2$.

\section*{Acknowledgments} 
\begin{acknowledgments}
The authors would like to thank H. Saito, and C. Hotta for valuable discussions. 
This work was supported by Grants-in-Aid for Scientific Research (KAKENHI Grant Nos. JP20KK0061, JP20H00130, JP21K18600, JP22H04933, JP22H01168, JP23H01124, JP23K22439, JP23K25821, JP23H04866, JP23H04870, JP23KJ1247 and JP24K00574) from the Japan Society for the Promotion of Science, by JST SPRING (Grant No. JPMJSP2110) from Japan Science and Technology Agency, by research support funding from the Kyoto University Foundation, by ISHIZUE 2024 of Kyoto University Research Development Program, and by Murata Science and Education Foundation.
In addition, liquid helium is supplied by the Low Temperature and Materials Sciences Division, Agency for Health, Safety and Environment, Kyoto University.
\end{acknowledgments}

\end{document}